\documentclass{aa}
\usepackage{graphicx}

\begin{document}

        \title{The 2007 outburst of the X-ray binary \object{XTE~J1856+053}
	\thanks{Based on observations obtained with XMM-Newton, an ESA science mission with 
	instruments and contributions directly funded by ESA Member States and NASA.}}

        \author{G.~Sala\thanks{\emph{Present address:} Dept. de F\'isica i Enginyeria Nuclear, EUETIB, Universitat Polit\`ecnica de Catalunya, c/ Compte d'Urgell 187, 08036 Barcelona, Spain. e-mail: gloria.sala@upc.edu}, 
		J.~Greiner, 
		M.~Ajello\thanks{\emph{Present address:} SLAC, 2575 Sand Hill Road, Menlo Park, CA 94025, and KIPAC, Stanford, CA 94305, USA} 
		\& N. Primak\thanks{\emph{Present address:} Skinakas Observatory, Physics Department, University of Crete, P.O. Box 2208, 71008 Heraklion, Crete, Greece}}

	\institute{Max-Planck-Institut f\"ur extraterrestrische Physik, Postfach 1312, D-85741 Garching, Germany}

        \offprints{G. Sala}

        \date{Received 31 March 2008 /Accepted 2 August 2008}

 
\abstract
	{}
	{On 28 February 2007 a new outburst of the previously known transient source 
XTE~J1856+053 was detected with RXTE/ASM. We present here the results of an XMM-Newton (0.5--10.0~keV) 
Target of Opportunity observation performed on 14 March 2007,
aimed at constraining the mass of the compact object in this X-ray binary and determining its main properties.}
	{The EPIC-pn camera was used in Timing mode and its spectrum fit together with the RGS data. 
IR observations with GROND at the 2.2~m telescope in La Silla provide further information on the system.}
	{The X-ray light curve shows that both the 1996 and the 2007 outbursts 
had two peaks. The X-ray spectrum is well fit with a thermal accretion disk model, 
with $kT_{in}=0.75(\pm0.01)$~keV and foreground absorption 
$N_{\rm{H}}=4.5(\pm0.1)\times10^{22}\,\mbox{cm}^{-2}$.
The low disk temperature favours a black-hole as accreting object, with an 
estimated mass in the range 1.3--4.2$M_{\odot}$. From the IR upper limits we argue 
that XTE~J1856+053 is a low mass X-ray binary. We estimate the
orbital period of the system to be between 3 and 12 hours.}
	{}


        \titlerunning{The 2007 outburst of  XTE~J1856+053}
        \authorrunning{Sala, Greiner, Ajello, Primak}
        \maketitle

%

\section{Introduction}
\label{intro}

X-ray binaries are the brightest X-ray sources in the sky. They are powered
by the accretion of material from the secondary star 
onto a compact object (neutron star or black hole). 
At present, around 20 X-ray binaries contain a dynamically confirmed black hole, and 
around another 20 are the so-called black hole candidates (Remillard \& McClintock~\cite{rmc06}). 
The spectral type of the companion determines the accretion regime and properties. In low mass X-ray binaries,
with secondary stars of type later than A, the mass transfer occurs through Roche lobe
overflow and forms an X-ray emitting accretion disk. 
The optical emission is dominated by the X-ray heated companion,
the outer disk and/or reprocessed hard X-rays. 
In the case of high mass X-ray binaries, where the companion is an O or B 
star, the strong wind of the secondary star is intercepted and accreted by the compact object.
The main X-ray source stems from the wind interaction. Though an accretion disk may also be present,
the secondary star is dominating the optical emission of the source.

The generally accepted picture for the X-ray emission of accreting black holes consists of 
an accretion disk, responsible for thermal black body emission in the X-ray band,
and a surrounding hot corona, the origin site of non-thermal power-law emission, 
up to the energy range of gamma-ray telescopes, due to inverse comptonization 
of soft X-ray photons from the disk. The standard accretion disk model, or multi-color disk model 
(Mitsuda et al. \cite{mit84}, \emph{diskbb} model in \emph{xspec}) 
consists of the sum of black bodies from an accretion disk with a surface temperature 
that depends on the radius $R$ as $R^{-3/4}$. This asymptotic distribution is sufficiently 
accurate for $R\gg6 R_{\rm{g}}$. But since the model neglects the torque-free boundary condition, 
the temperature distribution is not accurate for the inner disk when it extends down to the 
innermost stable orbit. The \emph{diskpn} model in \emph{xspec} (Gierli\'nski~et~al.~\cite{gie99})
includes corrections for the temperature distribution near the black hole 
by taking into account the torque-free inner-boundary condition.

Seven of the 20 confirmed black holes, and 12 of the black-hole candidates
are transient sources with only one unique outburst observed.
Here we summarize the main properties of XTE~J1856+053, 
an X-ray transient with two observed outbursts that may be the next in the black-hole candidate list.
XTE~J1856+053 was discovered with RXTE/PCA during a survey of the Galactic ridge 
in 1996 (Marshall~et~al.~\cite{mar96}). The RXTE/ASM light-curve of the 1996 outburst showed two peaks (Fig.~\ref{fig1}):
a first one starting on 1996 April 4 (MJD 50177), lasting for 27 days 
with symmetric rise and decline, and reaching 75~mCrab (2--12~keV);
and a second fast rise--slow decay (FRED) peak starting on 1996 Sept. 9 (MJD 50335), lasting for 70 days, and reaching 
a maximum flux of 79~mCrab (Remillard~\cite{rem99}).
This second maximum in X-rays was preceded 8 days before by a peak of 30--60~mCrab at higher energies (20--100~keV), 
detected by BATSE on 1996 Sept. 7--9 (MJD 50333--50335, Barret~et~al.~\cite{bar96}).

\section{The XTE~J1856+053 outburst in 2007}

\begin{figure*}
\centering
\resizebox{\hsize}{!}{\includegraphics[angle=0]{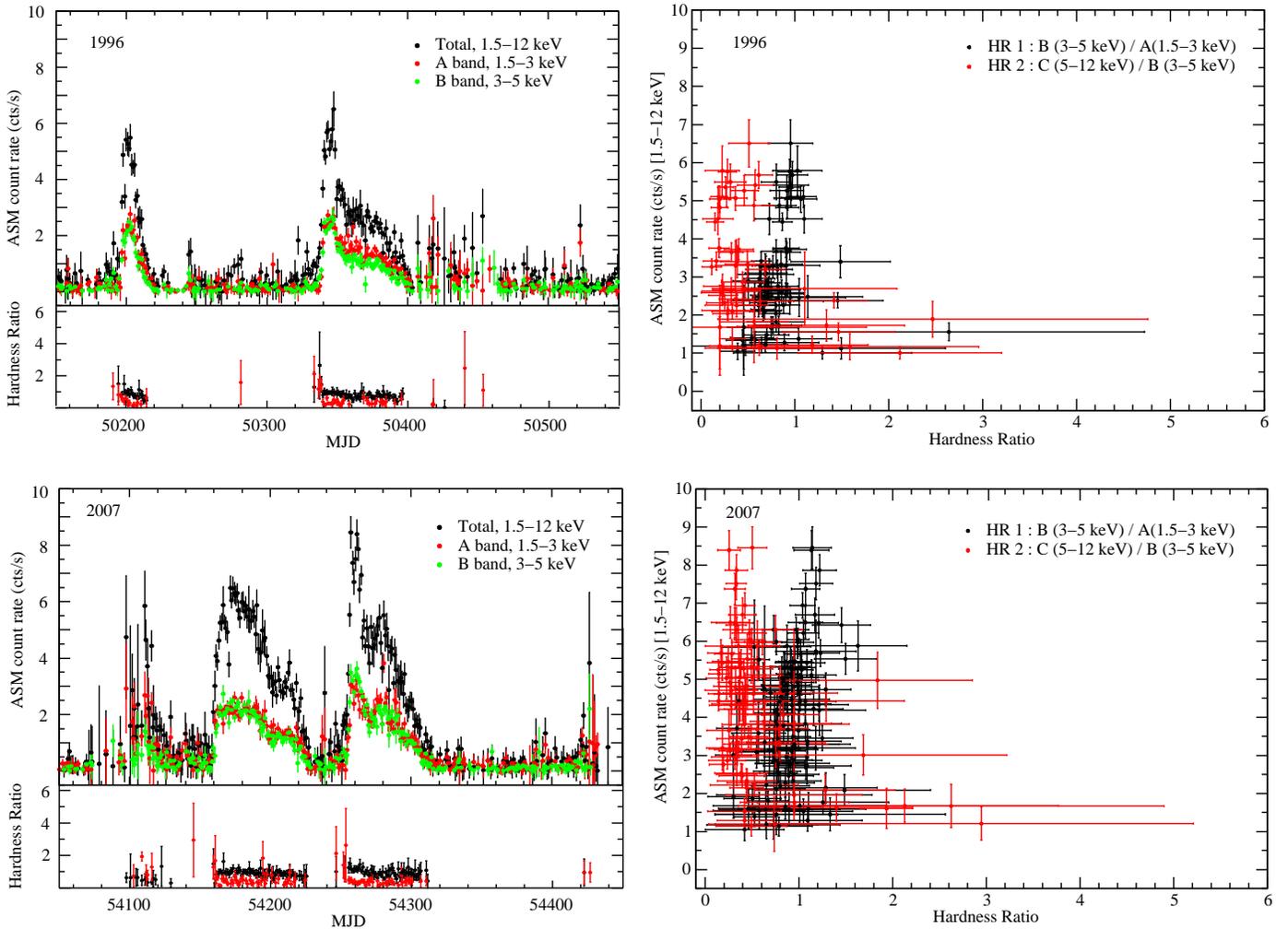}}
\caption{RXTE/ASM light curves and hardness ratios (left) and hardness ratio 
vs. intensity diagrams (HID, right) of XTE J1856+053 
during the outbursts in 1996 (top panels) and 2007 (bottom panels). 
Only hardness ratio data points with significant detections are plotted
(total count-rate larger than 1 c/s, hardness ratio smaller than 3, and error bars smaller than 100\%).
In the light curves, both the total count rate 
(1.5-12 keV, in black) and two bands (A: 1.5--3~keV in red, B: 3--5~keV in green) are shown.
In both the hardness ratio and the HID plots, HR1=B(3--5~keV)/A (1.5--3~keV) 
is shown in black and HR2=C(5--12~keV)/B(3--5~keV) is shown in red.}
\label{fig1}
\end{figure*}

In February 2007 a new outburst of XTE~J1856+053 was detected with RXTE (Levine~et~al.~\cite{lev07}).
As in the 1996 outburst, the RXTE/ASM light curve (1.5--12~keV) shows two peaks, 
preceded by a precursor on 2007 January 10--15 (MJD 54110--54115, Fig.~\ref{fig1}). 
The first peak started on 2007 February 28 (MJD~54159), 
reaching a maximum of $\sim$85~mCrab on  March 12 (MJD 54171) and lasting for $\sim$65~days. The second peak 
in 2007 started on May 21 (MJD 54241), rose in only 7 days to $\sim$110~mCrab, and lasted for $\sim$55~days.
As in the case of the second maximum in 1996, the two 2007 peaks were preceded by hard X-ray precursors 
detected by Swift/BAT (10--200~keV) in the periods February 22 to March 1 (MJD 54153--54160), 
and 2007 May 28--30 (MJD 54248--54250, Krimm~et~al.~\cite{kri07}).
As can be seen in the right panels in Fig.~\ref{fig1}, the hardness vs. intensity diagrams of both 
1996 and 2007 outbursts are also very similar.

The first peak of the 2007 outburst was followed up by Swift/XRT on March 10 (MJD~54169, Steeghs~et~al.~\cite{ste07}), 
RXTE/PCA on March 9 and 24 (MJD 54168 and 54183, Yamaoka~et~al.~\cite{yam07}), 
and Suzaku on March 19 (MJD 54178, Cackett~et~al.~\cite{cac07}).
In all cases the fit of the X-ray spectrum with multi-color disk models yielded inner disk temperatures 
in the range 0.7--0.9~keV, with foreground absorption column $N_H:(2.6-3.4)\times10^{22}\mbox{cm}^{-2}$.
No evidence for iron emission or absorption features was found. The RXTE/PCA spectra on March 9 and 24
were very similar, and they were fit with a multi-color disk model 
plus a power law, with photon index $\Gamma=2.0\pm0.5$ (Yamaoka~et~al.~\cite{yam07}).

Here we report the results of an XMM-Newton (0.5--10.0 keV) target of opportunity observation (TOO) requested 
shortly after the start of the 2007 outburst. 
The observation was executed on 14 March 2007 (MJD 54173), close to the maximum of the first peak.

\section{Results}
\label{sec_obs}

\subsection{XMM-Newton Target of Opportunity Observation}

We obtained a TOO observation of XTE~J1856+053 with XMM-Newton (Jansen~et~al.~\cite{jan01}) 
on 14 March 2007 (obs. ID. 0510010101, 5~ks for RGS, 1.5~ks for EPIC-pn), 
when the source flux detected by the RXTE/ASM was $\sim$80~mCrab (Sala~et~al.~\cite{sal07}). 
The observation was affected by solar flares, reducing the 
exposure time of the EPIC-pn camera to 1.5~ks of the 8~ks scheduled.
The EPIC-pn camera was used in Timing mode, 
and the RGSs in Spectroscopy HCR (High Count Rate) mode. 
The two MOS cameras were switched off for telemetry limitations. 
The Optical Monitor (OM) on board XMM-Newton 
obtained simultaneous images in the U band (six 1.5~ks exposures), 
but the source was not detected, with an upper limit for the AB magnitude in the U band of 23 mag.

XMM-Newton data were reduced with the XMM-Newton Science Analysis Software v7.1,
and XSPEC 11.3 was used for timing and spectral analysis. 
Following Kirsch~(\cite{kir07}) recommendations for the analysis of EPIC-pn data in Timing mode, 
only single and double events with energy higher than 0.5~keV were taken into account.
The source spectrum was extracted from a box including 10~RAWX columns at each side of 
the source position, and excluding events with RAWY~$\geq$~140 and RAWY~$\leq$~10.

\begin{figure}
\centering
\resizebox{\hsize}{!}{\includegraphics[angle=0]{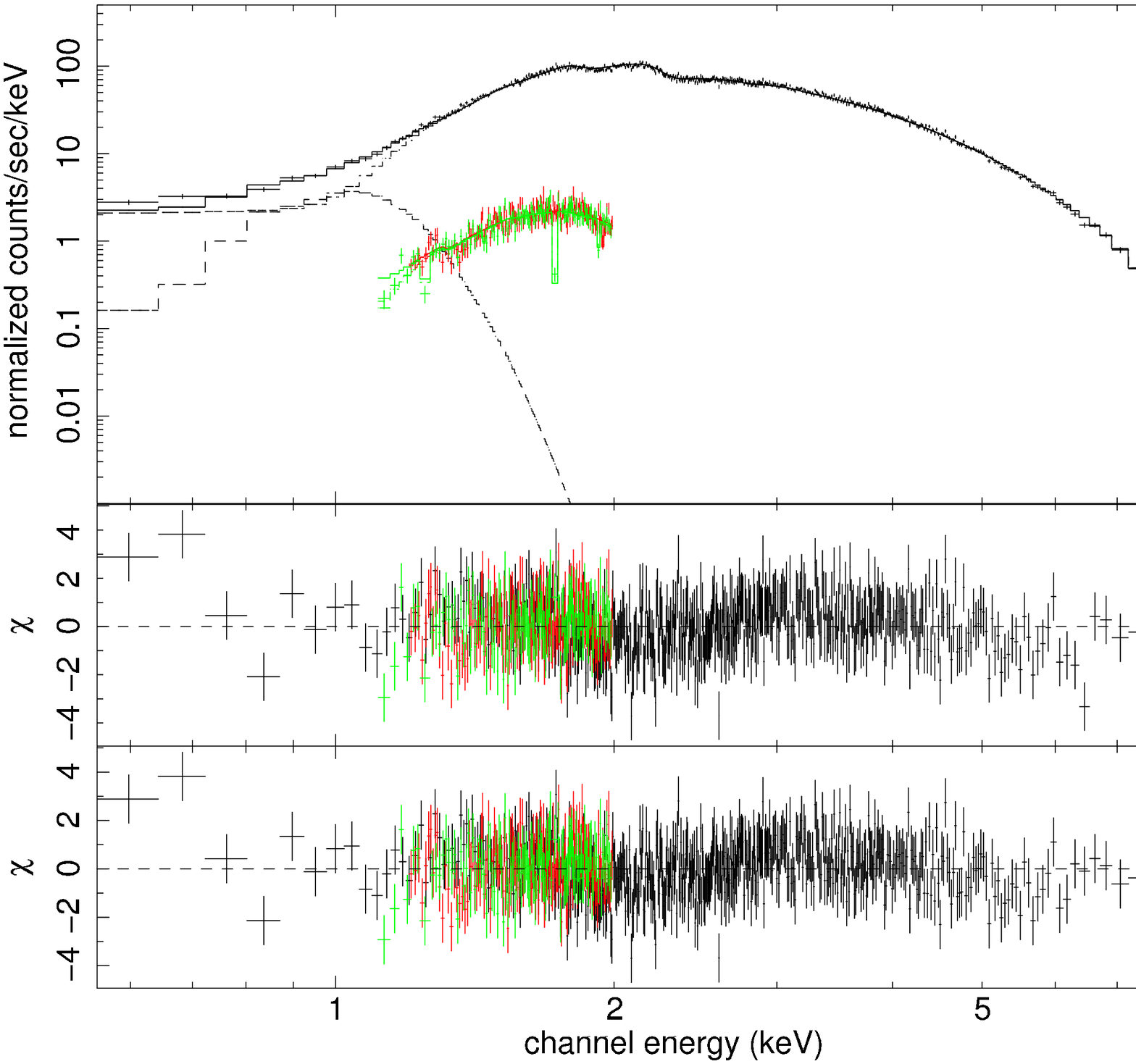}}
\caption{XMM-Newton X-ray spectra of XTE~J1856+053 on 14 March 2007. {\bf Top panel:}
EPIC-pn (black), RGS1 (red) and RGS2 (green) spectra, fitted with an 
absorbed multi-color accretion disk blackbody model plus a recombination edge at 0.87~keV.
{\bf Middle panel:} Residuals for the model used in upper panel. 
{\bf Bottom panel:} Residuals after adding an absorption line at 6.4~keV.}
\label{fig2}
\end{figure}

We fit simultaneously XMM-Newton EPIC-pn (0.5--10.0~keV), 
RGS1 and RGS2 (0.3--2.0~keV) data (Fig.~\ref{fig2}).
We fit the continuum spectrum with an absorbed multi-temperature
accretion disk blackbody model ({\it diskpn} model in {\it xspec}, Gierli\'nski et al.~\cite{gie99}).
We use the {\it tbabs} model for the absorption of the X-ray data, with Wilms~et~al.~(\cite{wil00}) 
abundances for the foreground absorption.

No indication of a hard component is evident in the residuals, 
but an excess is present below 1~keV, leading to a poor reduced 
$\chi^2$ of 1.99. Adding a recombination emission edge
at 0.87~keV (corresponding to O~VIII~K-shell) with
plasma temperature kT$=50(\pm3)$~eV improves the fit.
The significance of this feature is however to be taken with 
care, since the excess could be caused by some redistribution of 
higher energy photons to lower energies not properly 
taken into account by the calibration.
We have checked in addition that ignoring the pn data below 1~keV we 
obtain the same spectral results as adding the recombination edge at 0.87~keV. 
So this feature does not affect the results of the continuum spectrum.

The best fit to the continuum spectrum is obtained with 
$N_{\rm{H}}=4.5(\pm0.1)\times10^{22}\,\mbox{cm}^{-2}$, and
a disk with $kT_{\rm{in}}=0.75(\pm0.01)\,\mbox{keV}$. 
The relatively low temperature of the disk rules out a maximally spinning black hole, 
thus supporting the validity of the {\it diskpn} model for this fit.
A similar fit using the {\it diskbb} disk model yields a slightly higher inner temperature of 
$0.80(\pm0.01)$~keV.
The observed X-ray flux is 
$1.0(\pm0.1)\times10^{-9}\,\mbox{erg}\,\mbox{cm}^{-2}\,\mbox{s}^{-1}$ 
(0.5--10.0~keV), which corrected for absorption corresponds to an unabsorbed X-ray luminosity 
L$_{(0.5-10.0\,keV)}=4.0(\pm1.5)\times10^{38}({D}/10\rm{kpc})^2\,\mbox{erg}\,\mbox{s}^{-1}$.

With the above-mentioned best-fit model, the residuals in the EPIC-pn spectrum indicate the presence of a possible 
absorption line at 6.4 keV (bottom panel in Fig.~\ref{fig2}). 
The residuals improve with the addition of an absorption line with fix width 10~eV, at $6.4(\pm0.2)$~keV, 
with equivalent width $40(\pm35)$~eV (3$\sigma$ errors). The F-test probability of this component is 0.003.

We checked the upper limits for a possible hard component, related to the comptonization of the soft disk photons
by a hot corona. For a power law with photon index $\Gamma=2$ 
(or 3), we obtain an upper limit for the 5--10~keV flux of 
$9.7\times10^{-11}\mbox{erg}\,\mbox{cm}^{-2}\mbox{s}^{-1}$
($1.4\times10^{-12}\mbox{erg}\,\mbox{cm}^{-2}\mbox{s}^{-1}$ for $\Gamma=3$).

\subsection{Additional constraints in hard X-rays and IR from Swift/BAT and GROND}

We attempted to further constrain a possible hard component by using the Swift/BAT upper limit
for the period of the XMM-Newton observation. In particular, we extracted the data for the month of March 2007. 
Data processing was performed using standard Swift software available in the HEADAS 6.3.2
distribution and following the  recipes presented in Ajello~et~al.~(\cite{marco}).
Here, we recall the main data screening criteria:
(1) the star tracker must be locked and the spacecraft must be settled on the nominal pointing
direction,
(2) the spacecraft is outside the South Atlantic Anomaly and the total rate of the BAT array
is less than 18000~s$^{-1}$, and
(3) exposure must be longer than 200 s.
Data which fulfilled the above requirements were rebinned in energy accordingly to the
gain/offset map generated on board and were used to generate a sky image. The sky images were
searched for excesses above the 6-sigma threshold.
After the initial spike observed before the X-ray outburst (Krimm~et~al.~\cite{kri07}), 
the source was not detected by Swift/BAT. 
We obtain an upper limit of $10^{-10}\mbox{erg}\,\mbox{cm}^{-2}\mbox{s}^{-1}$ ($3\sigma$) in the 15--100~keV band.
This limit is at least two orders of magnitude larger than the flux of the brightest power-law component 
compatible with the EPIC-pn data, so no further constraint on the hard component can be obtained from the Swift/BAT data.

XTE~J1856+053 was observed on 26 August 2007 with GROND (Greiner~et~al.~\cite{grond}), 
a 7-channel imager mounted at the MPI/ESO 2.2m telescope at La Silla (Chile).
The source was not detected in a 8-minute exposure, with lower limits for the IR magnitudes
of J$\sim$20, H$\sim$19, and K$\sim$18 (calibrated against 2MASS).

\section{Discussion}
\label{sec_dis}

The presented X-ray spectrum of XTE~J1856+053 in March 2007, dominated by the 
accretion disk emission and without evidence for a power law component, corresponds 
to the thermal (high/soft) state of black hole binaries (Remillard \& McClintock~\cite{rmc06}).
From the best-fit parameters of the continuum spectrum, 
some constraints are obtained for the distance to the source and the mass of the compact object.

The observed absorbing hydrogen column is larger than the average galactic column density 
in the source direction, 1.4$\times10^{22}\rm{cm}^{-2}$ (Dickey \& Lockman~\cite{dic90}), which locates
XTE~J1856+053 behind the galactic slab of interstellar medium and 
provides a lower limit for the distance of 1~kpc.

The non-detection in the IR by GROND (with J$>$20) rules out the presence of a massive secondary in XTE~J1856+053. 
Even for a distance of 10~kpc, an O or B star would be detected with J$\sim$9--11. We can thus classify XTE~J1856+053 as a 
low mass X-ray binary (LMXB). We can also use the GROND non-detection to estimate a lower limit for the distance:
for a faint M2 secondary (the latest spectral type of a donor in a confirmed BH binary;
see Tab. 1 in McClintock \& Remillard \cite{mcrem04}), and with the extinction derived from
the X-ray absorption column (A$_J$=6.7, using $N_{\rm H}=5.9\times10^{21}E_{B-V}\,\mbox{cm}^{-2}$
and  A$_J$=0.282A$_V$; Zombeck \cite{zom07}, Rieke \& Liebofsky \cite{rl85})
the non-detection in the J band would set a lower limit for the distance of 1~kpc, 
consistent with the lower limit estimated from the absorption column in X-rays.

\begin{figure}
\centering
\resizebox{\hsize}{!}{\includegraphics[angle=0]{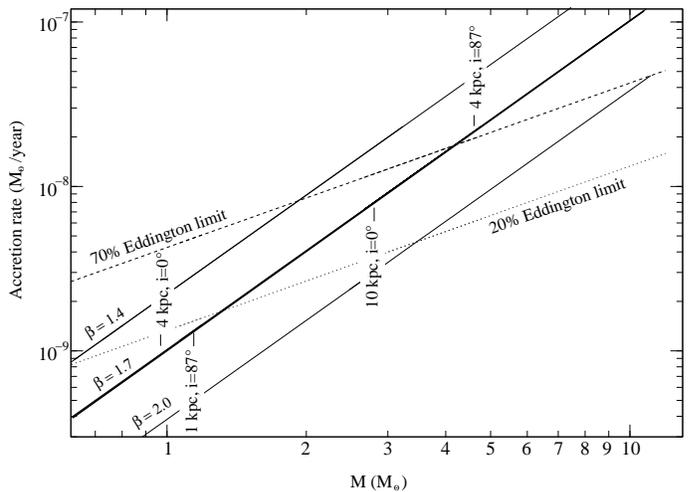}}
\caption{Accretion rate as a function of the black hole mass derived from the {\it dispkn} model. 
Solid lines are obtained from the best-fit $kT_{in}$ and normalization constant $K_{diskpn}$, 
giving values to the inclination and distance (values of some sample points are indicated), 
for three values of the color/effective temperature ratio $\beta$.
Overplotted are shown the accretion rate corresponding to 20\% and 70\% of the Eddington limit (see text for details).}
\label{figmm} 
\end{figure}

The low temperature of the accretion disk favours a black-hole as the 
accreting compact object. With the normalization constant {\it K} of the {\it diskpn} model 
we can put some constraint on the mass of the compact object {\it M}, as
we did previously for XTE~J1817-330 (Sala~et~al.~\cite{sal07xte1817}). 
The normalization constant {\it K} of the {\it diskpn} model is 
related to the mass of the compact object {\it M}, the 
distance to the source {\it D}, and the inclination of the disk {\it i} as $K=\frac{M^2 cos(i)}{D^2 \beta^4}$,
where $\beta$ is the color/effective temperature ratio. Furthermore, the accretion rate is related to
the mass of the compact object and the maximum temperature of the disk (Gierli\'nski et al.~\cite{gie99}). 
Assuming $\beta=1.7$ and with the best fit value for the normalization of the {\it diskpn} model 
($K_{\rm {diskpn}}=8.5(\pm0.4)\times10^{-3}\mbox{M}_{\odot}^2\mbox{kpc}^{-2}$), 
we plot in Fig.~\ref{figmm} the accretion rate as a function of the mass of the compact object, 
giving different values for inclinations and distances  (values are indicated for some sample points).

At the time of the XMM-Newton observation, the flux of the source 
was at the maximum of the first outburst detected in March 2007. However, a brighter
outburst was detected by RXTE/ASM in June 2007. 
At the time of our XMM-Newton observations, the flux was 70\% 
of the maximum detected in June 2007. 
Assuming the Eddington limit as the upper limit for the accretion rate at the 
maximum in June 2007, we plot the accretion rate 
corresponding to 70\% of the Eddington limit as an upper limit for the XMM-Newton results. 
For the minimum distance of 1~kpc, a minimum mass of 1.3~M$_{\odot}$ is obtained, for a high inclination system. 

The upper limit for the accretion rate at the time of the XMM-Newton observation
($0.7\rm M^{\rm{acc}}_{\rm{Edd}}$) sets an upper limit for the mass of the central object of 4.2~M$_{\odot}$, 
which confirms the possibility that XTE~J1856+053 hosts a stellar mass black hole.
We note, however, that this estimation depends on the color correction factor.
The value adopted here ($\beta=1.7$) is rather conservative in comparison, for example, 
to $\beta=2.6$ obtained by Shrader \& Titarchuk (\cite{shra03}). A larger color correction factor would, of course, 
imply a larger upper limit for the black hole mass. We also
note that some bright black hole transients reach super-Eddington luminosities at maximum. 
If this were the case for XTE J1856+053, the upper limit for the mass will be larger than 4.2 M$_{\odot}$
and still support the presence of a black hole in the system.

With our range for the central object mass (1.3--4.2~M$_{\odot}$) and 
assuming a spectral type K or M of the donor which fills its Roche lobe, 
we estimate the period of the system to between 3 and 12 hours. 
This rather short period makes XTE~J1856+053 an easy target for dynamical studies that would allow to establish the 
central object mass.

\begin{acknowledgements}
We thank Norbert Schartel and the XMM-Newton team for carrying out the 
TOO observation presented here. 
This work is based on observations obtained with XMM-Newton, an ESA science mission with 
instruments and contributions directly funded by ESA Member States and NASA, 
and supported by BMWI/DLR (FKZ 50 OX 0001) and the Max-Planck Society.
We acknowledge the RXTE/ASM team for making quick-look results available for public use. 
GS is supported through DLR (FKZ 50 OR 0405).
\end{acknowledgements}

\end{document}